\documentclass[preprint]{revtex4}
\usepackage[latin1]{inputenc}
\usepackage{graphicx,epstopdf}
\usepackage{amsmath}
\usepackage{amssymb}
\usepackage{subfigure}
\usepackage{epsfig}
\usepackage{bigints}
\usepackage[usenames]{color}

\begin{document}

\title{Waves in magnetized quark matter}

\author{D. A. Foga\c{c}a, S. M. Sanches Jr. and F. S. Navarra}

\address{Instituto de F\'isica, Universidade de S\~ao Paulo,
Rua do Mat\~ao Travessa R, 187, 05508-090 S\~ao Paulo, SP, Brazil}

\begin{abstract}

We study wave propagation in a non-relativistic cold  quark-gluon plasma
immersed in a constant magnetic field. Starting from the Euler equation we
derive linear wave equations and investigate their stability and causality.
We use a generic form for the equation of state, the EOS derived from the MIT
bag model and also a variant of the this model which includes gluon degrees of
freedom. The results of this analysis may be relevant for perturbations propagating through the quark matter
phase in the core of compact stars and also for perturbations propagating in the low
temperature quark-gluon plasma formed in low energy heavy ion collisions, to be carried out at FAIR
and NICA.

\end{abstract}

%\pacs{PACS Numbers : 12.39.Ba(Bag model),
%47.(Fluid dynamics), 12.38.-t(Quantum chromodynamics),
%12.38.Mh(Quark-gluon plasma)}

\maketitle

\section{Introduction}

There is a strong belief that quark gluon plasma (QGP) has been  formed in heavy
ion collisions at RHIC and at LHC \cite{qm,braun}. Deconfined quark matter may also exist in the
core of compact stars \cite{compa}. Waves may be formed in the QGP \cite{wave,we13,we15}.  In heavy ion collisions
waves may be produced, for example, by fluctuations in baryon number, energy density or temperature  caused by
inhomogeneous initial conditions \cite{fluc}.  Furthermore, there may be fluctuations induced by energetic partons,
which have been scattered in the initial collision of the two nuclei and propagate through the
medium, loosing energy and acting as a source term for the hydrodynamical equations.

In \cite{we13} we have studied wave propagation in cold and dense matter both in a
hadron gas phase and in a quark gluon plasma phase. In deriving wave equations from
the equations of hydrodynamics, we have considered both small and large
 amplitude
waves. The former were treated with the linearization approximation while the latter
were treated with the reductive perturbation method.
Linear waves were obtained by solving an inhomogeneous viscous wave equation and they
have the familiar form of sinusoidal traveling waves multiplied by an exponential
damping factor, which depends on the viscosity coefficients. Since these coefficients
differ by two orders of magnitude, even without any numerical calculation we  concluded
that, apart from extremely special parameter choices, in contrast to the quark gluon
plasma there will be no linear wave propagation in a hadron gas.

In this work we will investigate the  effects of a magnetic field on wave
propagation in a quark gluon plasma. We shall focus on the stability and causality of these
waves.  A natural question is  ``how does the magnetic field affect stability and causality of
density waves ?''.  We will try to answer this question in a, as much as possible, model
independent way.

Our conclusions should apply to the deconfined
cold quark matter in compact stars and to the cold (or slightly warm) quark gluon plasma
formed in heavy ion collisions at intermediate energies, to be performed at FAIR
\cite{fair} or NICA \cite{nica}.

In what follows we will carry out a wave analysis which is very frequently used in
hydrodynamcis \cite{instab}. We will be able to see if the presence of
a magnetic field modifies the conclusions reached in \cite{we13}.

\section{Hydrodynamics in an external magnetic field}

We shall consider the non-relativistic Euler equation \cite{land}
with an external magnetic field $\vec{B}$.
The three fermions species (three quarks) have negative or positive charges
and due to the external magnetic field they may assume different
trajectories \cite{azam,multif}. As a consequence we must apply the multifluid
approach \cite{azam,multif}, which consists in writing one Euler equation for each
quark $f=u,d,s$  :
\begin{equation}
{\rho_{m\,f}}\Bigg[{\frac{\partial \vec{v_f}}{\partial t}} +(\vec{v_f}\cdot \vec{\nabla})
\vec{v_f}\Bigg]=
-\vec{\nabla}p
+{\rho_{c\,f}}\Big(\vec{v_f} \times \vec{B} \Big)
\label{nsgeralmag}
\end{equation}
where ${\rho_{m\,f}}$  and  $\rho_{c\,f}$  are the  mass and charge density of the
quarks of flavor $f$ respectively.
We employ natural units ($\hbar=c=1$)
and the metric used is $g^{\mu\nu}=\textrm{diag}(+,-,-,-)$.

When we employ the multifluid approach, we are effectively using the approximation
of weak interactions between the fluid constituents.   In principle in an ideal QGP
the interaction between the quark and gluon constituents is weak. In the presence of
a strong magnetic field the interaction is even weaker, since the coupling constant
decreases with increasing B field \cite{gastao}. We will work with three equations of
state. In the first two of them there is no interaction between the constituents.
They are compatible with the
multifluid approach. In the third one (called ``mean field QCD'') we have interactions,
but the coupling
constant is not large. What justifies the mean field approximation is the
high density of sources. So we assume that in all our calculations we are in the weak
coupling regime and hence we can borrow all the techniques and approximations (including
the multifluid approach) from the plasmas known in electrodynamics.

In what follows  we will  consider quark matter  with three quark flavors: up ($u$),
down ($d$) and strange ($s$).  As it is usually studied in \cite{we16}, such quark matter may exist in compact stars.
The charges are: $Q_{u}= 2 \, Q_{e}/3$, $Q_{d}= - \, Q_{e}/3$  and $Q_{s}= - \, Q_{e}/3$,
where $Q_{e}=0.08542$ is the absolute value of the electron charge in
natural units \cite{we16,glend}.  The masses are \cite{quarkmassespdg}:
$m_{u}=2.2 \, MeV$, $m_{d}=4.7 \, MeV$ and $m_{s}=96 \, MeV$.

In the above equation  the pressure is a global feature of the fluid. The velocity,
masses and charges are specified for each fermion species. The equation of state
contains all fermions of the fluid under the
external magnetic field $\vec{B}$.  The magnetic field effects are included
both in the Euler equation and  in the equation of state.
We consider an uniform magnetic field of intensity $B$ in the $z-$direction described
by $\vec{B}=B \hat{z}$ .

The continuity equation for the baryon density ${\rho_{B}}_{f}$ reads \cite{land}:
\begin{equation}
{\frac{\partial {\rho_{B}}_{f}}{\partial t}} + {{\nabla}} \cdot ({\rho_{B}}_{f} \,
{\vec{v_f}})=0
\label{conteq}
\end{equation}
In general, the relationship between the mass density ($\rho_{m}$) and the particle
density ($\rho$) is given by $\rho_{m}=m\rho$, where $m$ is the particle mass. We
have then ${\rho_{m\,f}}=m_f \,\, {\rho}_{f}$ in (\ref{nsgeralmag}).  Besides, the quark number
density can be rewritten in terms of the respective baryon density as
${\rho_{m}}_f=3m_{f} \,\, {\rho_{B}}_{f}$, since
${\rho_{B}}_{f}=\rho_{f}/3$. The  charge density in (\ref{nsgeralmag}) of each quark is given
by ${\rho_{c}}_{u}=2Q_{e}\,{\rho_{B}}_{u}$\, ,
\, ${\rho_{c}}_{d}=-Q_{e}\,{\rho_{B}}_{d}$ \,\, and \,\,
${\rho_{c}}_{s}=-Q_{e}\,{\rho_{B}}_{s}$ . In short we have
${\rho_{c}}_{f}=3\,{Q_{f}} \, {\rho_{B}}_{f}$ for each quark $f$.

\section{Non-relativistic equation of state}

The equation of state of the quark gluon plasma can be written as:
\begin{equation}
p = {c_s}^2 \epsilon
\label{eosnr}
\end{equation}
where $p$, $\epsilon$ and $c_s$ are the pressure, energy density and
speed of sound respectively.
In the presence of an external magnetic field, we may have
two different pressures, one parallel ($p_{\parallel}$) and another perpendicular ($p_{\perp}$) to
the $B$ field direction. Consequently we will also have
a parallel (${c_s}_{\parallel}$) and a perpendicular (${c_s}_{\perp}$) speed of sound. They are
given by \cite{soundes} :
\begin{equation}
{({c_{s}}_{\parallel})}^{2}={\frac{\partial p_{\parallel}}{\partial \varepsilon}}
\hspace{1.0cm} \textrm{and} \hspace{1.0cm}
{({c_{s}}_{\perp})}^{2}={\frac{\partial p_{\perp}}{\partial \varepsilon}}
\label{soundes}
\end{equation}
and so $p_{\parallel} \approx {({c_{s}}_{\parallel})}^{2} \, \varepsilon$ and also $p_{\perp}
\approx {({c_{s}}_{\perp})}^{2} \, \varepsilon$.
In the non-relativistic limit we have \cite{we13}: $\varepsilon \cong {\rho_{m}}$ , where ${\rho_{m}}$
is the volumetric mass
density, which can be rewritten as
${\rho_{m}}=3m_{f}\,{\rho_{B}}_{f}$. Considering the pressure anisotropy  we have:
\begin{equation}
\vec{\nabla} p \approx  3m_{f}\Bigg({({c_{s}}_{\perp})}^{2}\,{\frac{\partial{{\rho_{B}}_{f}}}{\partial x}}\,,\,
{({c_{s}}_{\perp})}^{2}\,{\frac{\partial{{\rho_{B}}_{f}}}{\partial y}}\,,\,
{({c_{s}}_{\parallel})}^{2}\,{\frac{\partial{{\rho_{B}}_{f}}}{\partial z}}\hat{z}\Bigg)
\label{gradespresses}
\end{equation}
Inserting (\ref{gradespresses}) into (\ref{nsgeralmag}), we have for the ${f}$-quark:
\begin{equation}
3m_{f} \,\, {\rho_{B}}_{f}\Bigg[{\frac{\partial \vec{v}_{f}}{\partial t}} +(\vec{v}_{f}\cdot \vec{\nabla})\vec{v}_{f}\Bigg]=
$$
$$
-3m_{f}\Bigg({({c_{s}}_{\perp})}^{2}\,{\frac{\partial{{\rho_{B}}_{f}}}{\partial x}}\,,\,
{({c_{s}}_{\perp})}^{2}\,{\frac{\partial{{\rho_{B}}_{f}}}{\partial y}}\,,\,
{({c_{s}}_{\parallel})}^{2}\,{\frac{\partial{{\rho_{B}}_{f}}}{\partial z}}\Bigg)
+3\,{Q_{f}}\,\,{\rho_{B}}_{f}\Big(\vec{v}_{f} \times \vec{B} \Big)
\label{nsgeralmagffa}
\end{equation}

Linear waves are studied with the dispersion relation obtained through the linearization
formalism \cite{we13,we15}.  In this formalism  the Euler equation (\ref{nsgeralmagffa})
and the continuity equation (\ref{conteq}) are rewritten in terms
of the perturbed dimensionless variables for the densities, $\hat{\rho}_{B\,f}$, and also for the velocities, $\hat{\vec{v}}_{f}$,
defined from the equilibrium configuration (density $\rho_{0}$ and sound speed $c_{s}$).
The perturbations are described by the corresponding small deviations denoted by $\delta$:
\begin{equation}
\hat{\rho}_{B\,f}(\vec{x},t)={\frac{{\rho_{B}}_{f}(\vec{x},t)}{{\rho_{0}}}}=
1+\delta {\rho_{B}}_{f}(\vec{x},t)
\label{rhosadmlin}
\end{equation}
and
\begin{equation}
{\hat{\vec{v}}_{f}}(\vec{x},t)={\frac{\vec{v}_{f}(\vec{x},t)}{{c_{s}}}}=
\delta \vec{v}_{f}(\vec{x},t)
\label{velosadm}
\end{equation}
and only $\mathcal{O}(\delta)$ terms are considered. Inserting (\ref{rhosadmlin}) and
(\ref{velosadm}) into (\ref{nsgeralmagffa}) and into (\ref{conteq}), and linearizing both
equations, we find:
\begin{equation}
3m_{f} \,\, {\rho_{0}}\,{\frac{\partial }{\partial t}} \delta\tilde{\vec{v}}_{f}+3m_{f}\,\,{\rho_{0}}
\,\Bigg({({c_{s}}_{\perp})}^{2}\,{\frac{\partial}{\partial x}}\delta{\rho_{B}}_{f}\,,\,
{({c_{s}}_{\perp})}^{2}\,{\frac{\partial}{\partial y}}\delta{\rho_{B}}_{f}\,,\,
{({c_{s}}_{\parallel})}^{2}\,{\frac{\partial}{\partial z}}\delta{\rho_{B}}_{f}\Bigg)
$$
$$
-3\,{Q_{f}}\,\, {\rho_{0}}\,\Big(\delta\tilde{\vec{v}}_{f} \times \vec{B} \Big)=0
\label{nsgeralmagfagglin}
\end{equation}
and
\begin{equation}
{\frac{\partial}{\partial t}} {\delta\rho_{B}}_{f}+{\vec{\nabla}} \cdot \delta\tilde{\vec{v}}_{f}=0
\label{contfcompslin}
\end{equation}
where we have defined $\delta\tilde{\vec{v}}_{f}=\Big({c_{s}}_{\perp}\,
\delta {v_{f}}_{x}\,,\, {c_{s}}_{\perp}\,\delta {v_{f}}_{y}\,,\,{c_{s}}_{\parallel}\,
\delta {v_{f}}_{z}\Big)$ .

To study causality and stability, we follow the procedure adopted in
\cite{we13,instab, we15,moyses,shipu2,valis}, where the perturbations
are described by plane waves:
\begin{equation}
{\delta\varrho}={\mathcal{D}} \, e^{i\vec{k}\cdot \vec{x}-i\omega t}
\hspace{0.25cm} \textrm{,} \hspace{0.5cm}
\delta {{V}}_{x}={\mathcal{V}}_{x}\,\, e^{i\vec{k}\cdot \vec{x}-i\omega t}
\hspace{0.25cm} \textrm{,} \hspace{0.5cm}
\delta {{V}}_{y}={\mathcal{V}}_{y} \,\, e^{i\vec{k}\cdot \vec{x}-i\omega t}
\hspace{0.25cm} \textrm{and} \hspace{0.7cm}
\delta {{V}}_{z}={\mathcal{V}}_{z} \,\, e^{i\vec{k}\cdot \vec{x}-i\omega t}
\label{planewvs}
\end{equation}
with $\vec{k}\cdot \vec{x}=k_{x}\,x+k_{y}\,y+k_{z}\,z$ . The small amplitudes for the dimensionless variables are given by ${\mathcal{D}}$, ${\mathcal{V}}_{x}$, ${\mathcal{V}}_{y}$ and ${\mathcal{V}}_{z}$. In general, the frequency $\omega$ is decomposed as in \cite{we15,moyses,shipu2,valis}: $
\omega=\omega_{R}+i \omega_{I}$ with $\omega_{R} \in \mathbb{R}$ and $\omega_{I} \in \mathbb{R}$ .
Causality is ensured when the
following conditions for $\omega_R$ and $\omega_I$ are satisfied  \cite{japasbyref}:
\begin{equation}
\lim_{|\vec{k}| \to \infty} \Bigg| {\frac{\omega_{R}}{|\vec{k}|}} \Bigg| < 1
\label{japascondsa}
\end{equation}
and
\begin{equation}
\lim_{|\vec{k}| \to \infty} \Bigg| {\frac{\omega_{I}}{|\vec{k}|}} \Bigg| < \infty
\label{japascondsb}
\end{equation}
The condition  (\ref{japascondsa}) is equivalent to stating that the phase
velocity $|\vec{{{v}_{p}}}|$ is smaller than unity
(the speed of light in natural units) , i. e.  $| \vec{v_p} | < 1$, where
\begin{equation}
\vec{{{v}_{p}}}={\frac{\omega_{R}}{|\vec{k}|}}\hat{k}=
{\frac{\omega_{R}}{|\vec{k}|^{2}}}\vec{k}
\label{phasev}
\end{equation}
does not become greater as the wave number increases \cite{moyses,shipu2,valis}.
As a consistency check we evaluate
the group velocity, $({v}_{g})$, which is given by \cite{shipu2,valis,japasbyref}:
\begin{equation}
\vec{{{v}_{g}}}=\Bigg({\frac{\partial \omega_{R}}{\partial k_{x}}},
{\frac{\partial \omega_{R}\
}{\partial k_{y}}},{\frac{\partial
\omega_{R}}{\partial k_{z}}}  \Bigg)
\label{groupv}
\end{equation}
and must satisfy $|\vec{{{v}_{g}}}|< \infty$ as the wave number increases.  Stability is guaranteed when $\omega_{I}<0$\,, since
$e^{i\vec{k}\cdot \vec{x}-i\omega t}= e^{\omega_{I} t}e^{i\vec{k}\cdot \vec{x}-i\omega_{R} t}$
and $e^{\omega_{I} t}$ must be a decreasing function of  time.

Inserting (\ref{planewvs}) into the equations (\ref{nsgeralmagfagglin}) and (\ref{contfcompslin}), we are able to rewrite the
resulting equations in the following matrix form:
\begin{equation}
A(\omega,\vec{k}) \times
\left( \begin{array}{c}
{\delta\rho_{B}}_{f} \\
\delta {v_{f}}_{x} \\
\delta {v_{f}}_{y} \\
\delta {v_{f}}_{z}
\end{array} \right) = 0
\label{nreosgeralmatrix}
\end{equation}
where $A(\omega,\vec{k})$ is the matrix given by:
\begin{equation}
A(\omega,\vec{k})=
\left( \begin{array}{cccc}
i\,3m_{f}\,{\rho_{0}}\,({c_{s}}_{\perp})^{2}\,k_{x} & \hspace{0.3cm} -i\,3m_{f}\,{\rho_{0}}\,\omega\,({c_{s}}_{\perp}) &
\hspace{0.3cm} -3\,{Q_{f}}\,\, {\rho_{0}}\,B\,({c_{s}}_{\perp}) & \hspace{0.3cm} 0\\
i\,3m_{f}\,{\rho_{0}}\,({c_{s}}_{\perp})^{2}\,k_{y}  & \hspace{0.3cm} 3\,{Q_{f}}\,\, {\rho_{0}}\,B\,({c_{s}}_{\perp}) &
\hspace{0.3cm} -i\,3m_{f}\,{\rho_{0}}\,\omega\,({c_{s}}_{\perp}) & \hspace{0.3cm} 0\\
i\,3m_{f}\,{\rho_{0}}\,({c_{s}}_{\perp})^{2}\,k_{z}  & \hspace{0.3cm} 0 & \hspace{0.3cm} 0 &
\hspace{0.3cm} -i\,3m_{f}\,{\rho_{0}}\,\omega\,({c_{s}}_{\parallel}) \\
-i\,\omega  & \hspace{0.3cm} i\,({c_{s}}_{\perp})\,k_{x} & \hspace{0.3cm} i\,({c_{s}}_{\perp})\,k_{y}
& \hspace{0.3cm} i\,({c_{s}}_{\parallel})\,k_{z}
\end{array} \right)
\label{nreosAmatrix}
\end{equation}
The dispersion relation is found by solving the equation $det \,  A(\omega,\vec{k}) \, = \, 0$. It may be written as:
\begin{equation}
\omega^{4}-\Bigg[({c_{s}}_{\perp})^{2}\,{k_{x}}^{2}+({c_{s}}_{\perp})^{2}\,{k_{y}}^{2}+
({c_{s}}_{\parallel})^{2}\,{k_{z}}^{2}+\Bigg({\frac{B^{2}\,{Q_{f}}^{2}}{m_{f}^{2}}}\Bigg) \,\Bigg]\omega^{2}
+\Bigg({\frac{B^{2}\,{Q_{f}}^{2}}{m_{f}^{2}}}\Bigg)({c_{s}}_{\parallel})^{2}\,{k_{z}}^{2}=0
\label{nrdisprel}
\end{equation}
which implies that
\begin{equation}
{\omega^{2}}_{\pm}=
{\frac{({c_{s}}_{\perp})^{2}\,{k_{x}}^{2}}{2}}+{\frac{({c_{s}}_{\perp})^{2}\,{k_{y}}^{2}}{2}}+
{\frac{({c_{s}}_{\parallel})^{2}\,{k_{z}}^{2}}{2}}
+\Bigg({\frac{B^{2}\,{Q_{f}}^{2}}{2m_{f}^{2}}}\Bigg)
$$
$$
\pm\sqrt{{\frac{1}{4}}
\Bigg[({c_{s}}_{\perp})^{2}\,{k_{x}}^{2}+({c_{s}}_{\perp})^{2}\,{k_{y}}^{2}+
({c_{s}}_{\parallel})^{2}\,{k_{z}}^{2}+\Bigg({\frac{B^{2}\,{Q_{f}}^{2}}{m_{f}^{2}}}\Bigg) \,\Bigg]^{2}
-\Bigg({\frac{B^{2}\,{Q_{f}}^{2}}{m_{f}^{2}}}\Bigg)({c_{s}}_{\parallel})^{2}\,{k_{z}}^{2} }
\label{nrroots}
\end{equation}
The four solutions of (\ref{nrdisprel}) are then
$\omega(\vec{k})=\pm \sqrt{{\omega^{2}}_{\pm}}$ .  In this case we notice that $\omega(\vec{k})
\in \mathbb{R}$ and $\omega_{I}=0$ ensures stability.  The phase velocity is calculated from (\ref{phasev}):
\begin{equation}
\vec{{{v}_{p}}}={\frac{\omega}{|\vec{k}|}}\hat{k}=\pm \Bigg\{{\frac{({c_{s}}_{\perp})^{2}\,{k_{x}}^{2}}{2|\vec{k}|^{2}}}+
{\frac{({c_{s}}_{\perp})^{2}\,{k_{y}}^{2}}{2|\vec{k}|^{2}}}+
{\frac{({c_{s}}_{\parallel})^{2}\,{k_{z}}^{2}}{2|\vec{k}|^{2}}}
+\Bigg({\frac{B^{2}\,{Q_{f}}^{2}}{2m_{f}^{2}\,|\vec{k}|^{2}}}\Bigg)
$$
$$
\pm\sqrt{
\Bigg[{\frac{({c_{s}}_{\perp})^{2}\,{k_{x}}^{2}}{2|\vec{k}|^{2}}}+
{\frac{({c_{s}}_{\perp})^{2}\,{k_{y}}^{2}}{2|\vec{k}|^{2}}}+
{\frac{({c_{s}}_{\parallel})^{2}\,{k_{z}}^{2}}{2|\vec{k}|^{2}}}
+\Bigg({\frac{B^{2}\,{Q_{f}}^{2}}{2m_{f}^{2}\,|\vec{k}|^{2}}}\Bigg) \,\Bigg]^{2}
-\Bigg({\frac{B^{2}\,{Q_{f}}^{2}}{m_{f}^{2}}}\Bigg)({c_{s}}_{\parallel})^{2}\,
{\frac{{k_{z}}^{2}}{|\vec{k}|^{4}}} }
\,\, \Bigg\}^{1/2} \, \hat{k}
\label{nrphasev}
\end{equation}
With the above expression we can take the limit (\ref{japascondsa}):
\begin{equation}
\lim_{|\vec{k}| \to \infty} \Bigg| {\frac{\omega}{|\vec{k}|}} \Bigg|
\cong  \lim_{|\vec{k}| \to \infty}
\sqrt{
({c_{s}}_{\perp})^{2}
+ [({c_s}_{\parallel})^2 - ({c_s}_{\perp})^2] \,
\frac{k_z^2}{|\vec{k}|^2}
}
=
\sqrt{({c_{s}}_{\perp})^{2}
+ [({c_s}_{\parallel})^2 - ({c_s}_{\perp})^2] \, \cos^{2}(\theta)}
\label{nrjapascondsa}
\end{equation}
where $\theta$ is the angle between the direction of the magnetic field
and the direction of the wave propagation. We can see that the above limit
takes values between ${c_s}_{\parallel}$ and ${c_s}_{\perp}$.  Causality
is always satisfied.

The components of the group velocity (\ref{groupv}) are given by:
\begin{equation}
{\frac{\partial \omega}{\partial k_{x}}}=\pm{\frac{1}{2\omega}} \Bigg\{
({c_{s}}_{\perp})^{2}\,{k_{x}}
$$
$$
\pm {\frac{\Bigg[({c_{s}}_{\perp})^{2}|\vec{k}|^{2}-\Big[({c_{s}}_{\perp})^{2}-
({c_{s}}_{\parallel})^{2}\Big]{k_{z}}^{2}+ \Bigg({\frac{B^{2}\,{Q_{f}}^{2}}{m_{f}^{2}}}\Bigg)\Bigg]({c_{s}}_{\perp})^{2}\,{k_{x}}}
{2\,\sqrt{\Bigg[({c_{s}}_{\perp})^{2}{\frac{|\vec{k}|^{2}}{2}}-\Big[({c_{s}}_{\perp})^{2}-
({c_{s}}_{\parallel})^{2}\Big]{\frac{{k_{z}}^{2}}{2}}+ \Bigg({\frac{B^{2}\,{Q_{f}}^{2}}{2 m_{f}^{2}}}\Bigg)\Bigg]
-\Bigg({\frac{B^{2}\,{Q_{f}}^{2}}{m_{f}^{2}}}\Bigg)({c_{s}}_{\parallel})^{2}
\,{k_{z}}^{2}}}} \,\, \Bigg\}
\label{nrxgroupv}
\end{equation}
\begin{equation}
{\frac{\partial \omega}{\partial k_{y}}}=\pm{\frac{1}{2\omega}} \Bigg\{
({c_{s}}_{\perp})^{2}\,{k_{y}}
$$
$$
\pm {\frac{\Bigg[({c_{s}}_{\perp})^{2}|\vec{k}|^{2}-\Big[({c_{s}}_{\perp})^{2}-
({c_{s}}_{\parallel})^{2}\Big]{k_{z}}^{2}+ \Bigg({\frac{B^{2}\,{Q_{f}}^{2}}{m_{f}^{2}}}\Bigg)\Bigg]({c_{s}}_{\perp})^{2}\,{k_{y}}}
{2\,\sqrt{\Bigg[({c_{s}}_{\perp})^{2}{\frac{|\vec{k}|^{2}}{2}}-\Big[({c_{s}}_{\perp})^{2}-
({c_{s}}_{\parallel})^{2}\Big]{\frac{{k_{z}}^{2}}{2}}+ \Bigg({\frac{B^{2}\,{Q_{f}}^{2}}{2 m_{f}^{2}}}\Bigg)\Bigg]
-\Bigg({\frac{B^{2}\,{Q_{f}}^{2}}{m_{f}^{2}}}\Bigg)({c_{s}}_{\parallel})^{2}
\,{k_{z}}^{2}}}} \,\, \Bigg\}
\label{nrygroupv}
\end{equation}
and
\begin{equation}
{\frac{\partial \omega}{\partial k_{z}}}=\pm{\frac{1}{2\omega}} \Bigg\{
({c_{s}}_{\parallel})^{2}\,{k_{z}}
$$
$$
\pm {\frac{\Bigg[({c_{s}}_{\perp})^{2}|\vec{k}|^{2}-\Big[({c_{s}}_{\perp})^{2}-
({c_{s}}_{\parallel})^{2}\Big]{k_{z}}^{2}+ \Bigg({\frac{B^{2}\,{Q_{f}}^{2}}{m_{f}^{2}}}\Bigg)\Bigg]({c_{s}}_{\parallel})^{2}\,{k_{z}}
-\Bigg({\frac{2B^{2}\,{Q_{f}}^{2}}{m_{f}^{2}}}\Bigg)({c_{s}}_{\parallel})^{2}
\,{k_{z}}}
{2\,\sqrt{\Bigg[({c_{s}}_{\perp})^{2}{\frac{|\vec{k}|^{2}}{2}}-\Big[({c_{s}}_{\perp})^{2}-
({c_{s}}_{\parallel})^{2}\Big]{\frac{{k_{z}}^{2}}{2}}+ \Bigg({\frac{B^{2}\,{Q_{f}}^{2}}{2 m_{f}^{2}}}\Bigg)\Bigg]
-\Bigg({\frac{B^{2}\,{Q_{f}}^{2}}{m_{f}^{2}}}\Bigg)({c_{s}}_{\parallel})^{2}
\,{k_{z}}^{2}}}} \,\, \Bigg\}
\label{nrzgroupv}
\end{equation}
where we verify that $|\vec{{{v}_{g}}}|< \infty$ as the wave number increases.
From the results given by (\ref{nrjapascondsa}) and from the limit $\lim_{|\vec{k}| \to \infty} |\vec{{{v}_{g}}}|< \infty$
we conclude that causality is satisfied. Two particular cases have special interest:
\vskip0.5cm
\noindent
$(i)$  {\it No B field} (${c_{s}}_{\parallel} =  {c_{s}}_{\perp} =  c_{s} $): \,\,\, $\omega(\vec{k})=\pm (c_{s}) |\vec{k}|$ \,\,\,\, and
\,\,\,\, $|\vec{{{v}_{p}}}|= c_{s}$

\vskip0.7cm
\noindent
$(ii)$ {\it Very strong  B field} ($B^{2} \to \infty$): \,\,\, $
\omega(\vec{k}) \cong \, \pm \Bigg({\frac{B\,{Q_{f}}}{{m_{f}}}}\Bigg)$ \,\,\,\, and
\,\,\,\, $|\vec{{{v}_{p}}}|= \Bigg({\frac{B\,{Q_{f}}}{{m_{f}}|\vec{k}|}}\Bigg)$

\vspace{0.8cm}

In order to have an idea of the numbers involved, we remember that the relevant
strong magnetic fields are of the order of (or smaller than) $10^{19}$ $G$. These  values correspond to $B Q_e \simeq m_{\pi}^2
\simeq 0.02\, GeV^{2}$, with $m_{\pi} \simeq 140 \, MeV$, $1 \, GeV^{2}=1.44
\times 10^{19} \, G$ \cite{CAxe} and to a phase velocity of
\begin{equation}
|\vec{{{v}_{p}}}| = v_p \simeq \frac{B \,Q_e}{ m_f \, |\vec{k}|} \simeq
\frac{m^2_{\pi}}{ m_f \, |\vec{k}|}
\label{giantB}
\end{equation}
and hence $|\vec{{{v}_{p}}}|  < 1$ when $|\vec{k}| > 1000$ MeV, for example.

The above results for the non-relativistic equation of state are model independent
and allow for quantitative estimates of some quantities, as long as we stay far
from the very high velocity regime.

\section{The MIT Bag Model equation of state}

The thermodynamical  properties of the hot QGP can be calculated from first
principles in
lattice QCD. On the other hand, the equation of state of the cold quark gluon
plasma  is
not yet known with the same  level of precision and we need to use models. For
simplicity
we often use the equation of state derived from the MIT bag model, which
describes a gas of noninteracting quarks and gluons and takes into account
non-perturbative
effects through the bag constant $\mathcal{B}$. This constant is interpreted as the energy
needed to create a bubble (or bag) in the QCD physical vacuum. In our case the quarks move
under the action of  an external magnetic field.

The energy density ($\varepsilon_{MIT}$), the parallel pressure (${p_{f\,\parallel}}_{MIT}$)
and the perpendicular pressure (${p_{f\,\perp}}_{MIT}$), are given respectively by \cite{mitmags}:
\begin{equation}
\varepsilon_{MIT}={\mathcal{B}}+{\frac{B^{2}}{8\pi}}
+\sum_{f=u}^{d,s}{\frac{|Q_{f}|B}{2\pi^{2}}}\sum_{n=0}^{n^{f}_{max}} 3(2-\delta_{n0})
\int_{0}^{k^{f}_{z,F}} dk_{z}\sqrt{m_{f}^{2}+k_{z}^{2}+2n|Q_{f}|B}
\label{epsilontempzeromagonMIT}
\end{equation}
\begin{equation}
{p_{\parallel}}_{MIT}=-{\mathcal{B}}-{\frac{B^{2}}{8\pi}}
+\sum_{f=u}^{d,s}{\frac{|Q_{f}|B}{2\pi^{2}}}  \sum_{n=0}^{n^{f}_{max}} 3(2-\delta_{n0})
\int_{0}^{k^{f}_{z,F}} dk_{z} \,{\frac{{k_{z}}^{2}}{\sqrt{m_{f}^{2}+k_{z}^{2}+2n|Q_{f}|B}}}
\label{parallelpressuremagonMIT}
\end{equation}
\begin{equation}
{p_{\perp}}_{MIT}=-{\mathcal{B}}+{\frac{B^{2}}{8\pi}}
+\sum_{f=u}^{d,s}{\frac{|Q_{f}|^{2}B^{2}}{2\pi^{2}}}
 \sum_{n=0}^{n^{f}_{max}} 3(2-\delta_{n0}) n \int_{0}^{k^{f}_{z,F}}
{\frac{dk_{z}}{\sqrt{m_{f}^{2}+k_{z}^{2}+2n|Q_{f}|B}}}
\label{perppressuremagonMIT}
\end{equation}
The baryon density ($\rho_{B}$) is written as:
\begin{equation}
\rho_{B}=\sum_{f=u}^{d,s}\,{\frac{|Q_{f}|B}{2\pi^{2}}} \sum_{n=0}^{n^{f}_{max}}(2-\delta_{n0}) \, k^{f}_{z,F}(n)
\hspace{0.7cm} \textrm{with} \hspace{0.4cm}
n\leq n^{f}_{max}=int\Bigg[{\frac{{\mu_{f}}^{2}-m_{f}^{2}}{2|Q_{f}|B}}\Bigg]
\label{magbaryondensMIT}
\end{equation}
where the Fermi momentum is given by:
\begin{equation}
k^{f}_{z,F}(n)=\sqrt{{\mu_{f}}^{2}-m_{f}^{2}-2n|Q_{f}|B} ,
\label{kff}
\end{equation}
where $\mu_{f}$ is the chemical potential of the quark $f$ and $ {\it{int}}[a]$ denotes the integer part of $a$.
The parallel and perpendicular speed of sound in this case are given by (\ref{soundes}) :
${({c_{s}}_{\parallel})}^{2}= \partial {p_{\parallel}}_{MIT}/ \varepsilon_{MIT}$
\, and \, ${({c_{s}}_{\perp})}^{2}= \partial {p_{\perp}}_{MIT}/ \varepsilon_{MIT}$ \, .

In order to appreciate more easily the effect of the magnetic field,
we will consider the particular case of a very strong field, i.e., we
consider $|Q_{f}|B \, > \, {\mu_{f}}^{2}$ such that $n^{f}_{max}=0$
in (\ref{magbaryondensMIT}).  We choose a common chemical potential
$\mu$ which satisfies $|Q_{f}|B \, > \, \mu^{2} \, > \, m_{f}^{2}$ for all
quark flavors and  defines the following Fermi momentum:
$k^{f}_{z,F}(n)\rightarrow  k_{F}=\mu$. The baryon density
(\ref{magbaryondensMIT}) is then given by:
\begin{equation}
\rho_{B}=\sum_{f=u}^{d,s}\,{\frac{|Q_{f}|B}{2\pi^{2}}}\,\mu
\label{magbaryondensaltosB}
\end{equation}
In this limit the energy density and the pressures are given by
(\ref{epsilontempzeromagonMIT}),
(\ref{parallelpressuremagonMIT}) and (\ref{perppressuremagonMIT}):
\begin{equation}
\varepsilon_{MIT}={\mathcal{B}}+{\frac{B^{2}}{8\pi}}
+\sum_{f=u}^{d,s}{\frac{3|Q_{f}|B}{2\pi^{2}}}
\Bigg[-{\frac{m_{f}^{2}}{4}} \ln\Big(m_{f}^{2}\Big)
+{\frac{m_{f}^{2}}{2}}\ln\Big(2\,k_{F}\Big)+
{\frac{k_{F}^{2}}{2}} \Bigg)\Bigg]
\label{epsilontempzeromagonmitaltosB}
\end{equation}
\begin{equation}
{p_{\parallel}}_{MIT}=-{\mathcal{B}}-{\frac{B^{2}}{8\pi}}
+\sum_{f=u}^{d,s}{\frac{3|Q_{f}|B}{2\pi^{2}}}
\Bigg[{\frac{m_{f}^{2}}{4}} \ln\Big(m_{f}^{2}\Big)
-{\frac{m_{f}^{2}}{2}}\ln\Big(2\,k_{F}\Big)+
{\frac{k_{F}^{2}}{2}} \Bigg)\Bigg]
\label{parallelpressuremagonmitaltosB}
\end{equation}
\begin{equation}
{p_{\perp}}_{MIT}=-{\mathcal{B}}+{\frac{B^{2}}{8\pi}}
\label{perppressuremagonmitaltosB}
\end{equation}
Using the above expressions, the  pressure gradient is given by:
\begin{equation}
\vec{\nabla}{p}=\Bigg({\frac{\partial}{\partial x}}\,
{p_{\perp}}\,,\,{\frac{\partial }{\partial y}}\,{p_{\perp}}\,,\,
{\frac{\partial }{\partial z}}\,{p_{\parallel}}\Bigg)
=\Bigg(0 \,,\,0\,,\,-{\frac{3|Q_{f}|B\,m_{f}^{2}}{4\pi^{2}\,{\rho_{B}}_{f}}}{\frac{\partial {\rho_{B}}_{f}}{\partial z}}+
{\frac{6\pi^{2}}{|Q_{f}|B}}\,{\rho_{B}}_{f}{\frac{\partial {\rho_{B}}_{f}}{\partial z}} \Bigg)
\label{gradpresscartsonlyquarksstrongB}
\end{equation}
Repeating the same calculations of the last sections, the matrix
$A(\omega,\vec{k})$ in this case is:
\begin{equation}
A(\omega,\vec{k})=
\left( \begin{array}{cccc}
0 & \hspace{0.3cm} -i\,3m_{f}\,{\rho_{0}}\,\omega\,({c_{s}}_{\perp}) &
\hspace{0.3cm} -3\,{Q_{f}}\,\, {\rho_{0}}\,B\,({c_{s}}_{\perp}) & \hspace{0.3cm} 0\\
0  & \hspace{0.3cm} 3\,{Q_{f}}\,\, {\rho_{0}}\,B\,({c_{s}}_{\perp}) &
\hspace{0.3cm} -i\,3m_{f}\,{\rho_{0}}\,\omega\,({c_{s}}_{\perp}) & \hspace{0.3cm} 0\\
i\,\Omega_{s}\,k_{z}  & \hspace{0.3cm} 0 & \hspace{0.3cm} 0 &
\hspace{0.3cm} -i\,3m_{f}\,{\rho_{0}}\,\omega\,({c_{s}}_{\parallel}) \\
-i\,\omega  & \hspace{0.3cm} i\,({c_{s}}_{\perp})\,k_{x} & \hspace{0.3cm} i\,({c_{s}}_{\perp})\,k_{y}
& \hspace{0.3cm} i\,({c_{s}}_{\parallel})\,k_{z}
\end{array} \right)
\label{smiteosAmatrix}
\end{equation}
where
\begin{equation}
\Omega_{s} \equiv \Bigg({\frac{6\pi^{2}\,{\rho_{0}}^{2}}{|Q_{f}|B}}
-{\frac{3|Q_{f}|B\,m_{f}^{2}}{4\pi^{2}}} \Bigg)
\label{OmsmiteosAmatrix}
\end{equation}
and the dispersion relation is:
\begin{equation}
\omega^{4}-\Bigg[({\mathcal{V}}_{s})^{2} \,{k_{z}}^{2}+
\Bigg({\frac{B^{2}\,{Q_{f}}^{2}}{m_{f}^{2}}}\Bigg) \,\Bigg]\omega^{2}
+\Bigg({\frac{B^{2}\,{Q_{f}}^{2}}{m_{f}^{2}}}\Bigg)
({\mathcal{V}}_{s})^{2} \,{k_{z}}^{2}=0
\label{smitdisprelA}
\end{equation}
with the parameter ${\mathcal{V}}_{s}$ identified as:
\begin{equation}
({\mathcal{V}}_{s})^{2} \equiv {\frac{2\pi^{2}\,{\rho_{0}}}{|Q_{f}|B\,m_{f}}} -
{\frac{|Q_{f}|B\,m_{f}}{4\pi^{2}\,{\rho_{0}}}}
\label{newsoundstrongB}
\end{equation}
Considering (\ref{magbaryondensaltosB}) as the background density, we can
rewrite (\ref{newsoundstrongB}) as:
\begin{equation}
({\mathcal{V}}_{s})^{2} ={\frac{\tilde{Q}\,\mu}{|Q_{f}|\,m_{f}}} -
{\frac{|Q_{f}|\,m_{f}}{2\,\tilde{Q}\,\mu}}
\label{newsoundstrongBdetails}
\end{equation}
where $\tilde{Q}\equiv\sum_{j=u}^{d,s}\,|Q_{f}|$.  We clearly notice in
(\ref{newsoundstrongBdetails}) that $({\mathcal{V}}_{s})^{2}>0$ because
$\tilde{Q}>|Q_{f}|$ and $\mu > m_{f}$.  Inserting the above expression into
(\ref{smitdisprelA}) we can solve it, finding  $\omega$ and then the
phase and group velocities. The resulting expressions coincide with
equations (\ref{nrroots}) to (\ref{nrzgroupv}), once we set in these latter
${c_{s}}_{\perp}=0$ and  ${c_{s}}_{\parallel} \to {\mathcal{V}}_{s}$.
The  dispersion relation (\ref{smitdisprelA}) has
only real roots $(\omega_I=0)$ and always satisfies the stability condition
(\ref{japascondsb}). In particular, the new version of eq.
(\ref{nrjapascondsa}) is:
\begin{equation}
\lim_{|\vec{k}| \to \infty} \Bigg| {\frac{\omega}{|\vec{k}|}} \Bigg|
= \lim_{|\vec{k}| \to \infty} |\vec{{{v}_{p}}}| \cong \lim_{|\vec{k}| \to \infty} \sqrt{{\frac{
({\mathcal{V}}_{s})^{2}\,{k_{z}}^{2}}{|\vec{k}|^{2}}} }
= {\mathcal{V}}_{s} \,  \cos (\theta)
\label{sBmitjapascondsaA}
\end{equation}
where $\theta$ is, as before, the angle between the
vector $\vec{k}$ and
the $z$ direction.  Since ${\mathcal{V}}_{s}$ is always larger than one,
causality is guaranteed only for certain directions of propagation.
Perturbations propagating along the direction of the magnetic field
(for which $\theta = 0$ and $k_{z} = |\vec{k}|$), will have
$|\vec{{{v}_{p}}}| > 1$. This is unphysical and is an indication of the
inadequacy of the formalism for these extreme conditions.

\section{Improved  MIT  Bag  Model}

In this section we shall use the equation of state which we call mQCD and which was derived in
\cite{we11,we16}.  With mQCD we improve the MIT bag model including explicitly the gluonic  degrees
of freedom and  also new
non-perturbative effects. We assume that the quarks and gluons in the cold QGP are deconfined but can
interact,
forming the QGP. This means that the
coupling is nonzero and also that there are remaining non-perturbative interactions and
gluon condensates. We split the gluon field  into two components
$G^{a\mu}={A}^{a\mu}+{\alpha}^{a\mu}$, where ${A}^{a\mu}$ (``soft'' gluons) and
${\alpha}^{a\mu}$ (``hard''gluons) are the components of the field associated with low and
high momentum modes respectively. The expectation values of
${A}^{a\mu}{A}^{a}_{\mu}$ and ${A}^{a\mu}{A}^{a}_{\mu} {A}^{b\nu}{A}^{b}_{\nu}$ are
non-vanishing in a non-trivial vacuum and from them we  obtain
an effective gluon mass ($m_{G}$)
and also a contribution (${\mathcal{B}}_{QCD}$) to the energy and to the pressure  of the
system similar to the one of the MIT bag model. Since the number
of quarks is very large and their coupling to the gluons is not small, the high momentum
levels of the gluon field will have large occupation
numbers and hence the ${\alpha}^{a\mu}$ component of the field can be approximated by a
classical field. This is the same mean field approximation
very often applied to  models of nuclear matter, such as the Walecka model
\cite{mft,weset,we13}.

The energy density ($\varepsilon$), the parallel pressure ($p_{f\,\parallel}$) and the perpendicular
pressure ($p_{f\,\perp}$), are given respectively by
\cite{we16}:
\begin{equation}
\varepsilon={\frac{27{g_{h}}^{2}}{16{m_{G}}^{2}}}\,({{\rho_{B}}})^{2}+{\mathcal{B}}_{QCD}+
{\frac{B^{2}}{8\pi}}
+\sum_{f=u}^{d,s}{\frac{|Q_{f}|B}{2\pi^{2}}}\sum_{n=0}^{n^{f}_{max}} 3(2-\delta_{n0})
\int_{0}^{k^{f}_{z,F}} dk_{z}\sqrt{m_{f}^{2}+k_{z}^{2}+2n|Q_{f}|B}
\label{epsilontempzeromagon}
\end{equation}
\begin{equation}
p_{\parallel}={\frac{27{g_{h}}^{2}}{16{m_{G}}^{2}}}\,({{\rho_{B}}})^{2}
-{\mathcal{B}}_{QCD}-{\frac{B^{2}}{8\pi}}
+\sum_{f=u}^{d,s}{\frac{|Q_{f}|B}{2\pi^{2}}}  \sum_{n=0}^{n^{f}_{max}} 3(2-\delta_{n0})
\int_{0}^{k^{f}_{z,F}} dk_{z} \,{\frac{{k_{z}}^{2}}{\sqrt{m_{f}^{2}+k_{z}^{2}+2n|Q_{f}|B}}}
\label{parallelpressuremagon}
\end{equation}
\begin{equation}
p_{\perp}={\frac{27{g_{h}}^{2}}{16{m_{G}}^{2}}}\,({{\rho_{B}}})^{2}
-{\mathcal{B}}_{QCD}+{\frac{B^{2}}{8\pi}}
+\sum_{f=u}^{d,s}{\frac{|Q_{f}|^{2}B^{2}}{2\pi^{2}}}  \sum_{n=0}^{n^{f}_{max}} 3(2-\delta_{n0}) n
\int_{0}^{k^{f}_{z,F}}   {\frac{dk_{z}}{\sqrt{m_{f}^{2}+k_{z}^{2}+2n|Q_{f}|B}}}
\label{perppressuremagon}
\end{equation}
The baryon density ($\rho_{B}$) is given by (\ref{magbaryondensMIT}) \cite{we16}.

As in \cite{we16,we12} we  define $\xi \equiv g_{h}/m_{G}$. Choosing $\xi=0$ we recover the MIT
EOS (\ref{epsilontempzeromagonMIT}),
(\ref{parallelpressuremagonMIT}) and (\ref{perppressuremagonMIT}). For a given magnetic field intensity,
we choose the values for the
chemical potentials ${\nu_{f}}$  which  determine the density $\rho_{B}$.
We also choose the other parameters:
$\xi$ and ${\mathcal{B}}_{QCD}$.  The background density  (upon which  small perturbation occur) is given by $\rho_{0}$, and it is usually given as multiples of the ordinary nuclear matter density $\rho_{N}=0.17\, fm^{-3}$ \cite{we16}.

Performing the same calculations shown in the previous sections, we obtain the
following matrix:
\begin{equation}
A(\omega,\vec{k})=
\left( \begin{array}{cccc}
i\Bigg({\frac{27\,{g_{h}}^{2}\,{\rho_{0}}^{2}}{8\,{m_{G}}^{2}}}\Bigg)\,k_{x} & \hspace{0.3cm}
\vspace{0.2cm} -i\,3m_{f}\,{\rho_{0}}\,\omega\,({c_{s}}_{\perp}) &
\hspace{0.3cm} -3\,{Q_{f}}\,\, {\rho_{0}}\,B\,({c_{s}}_{\perp}) & \hspace{0.3cm} 0\\ \vspace{0.2cm}
i\Bigg({\frac{27\,{g_{h}}^{2}\,{\rho_{0}}^{2}}{8\,{m_{G}}^{2}}}\Bigg)\,k_{y}  & \hspace{0.3cm} 3\,{Q_{f}}\,\, {\rho_{0}}\,B\,({c_{s}}_{\perp}) &
\hspace{0.3cm} -i\,3m_{f}\,{\rho_{0}}\,\omega\,({c_{s}}_{\perp}) & \hspace{0.3cm} 0\\
\vspace{0.2cm}
i\Bigg({\frac{27\,{g_{h}}^{2}\,{\rho_{0}}^{2}}{8\,{m_{G}}^{2}}}\Bigg)\,k_{z}  & \hspace{0.3cm} 0 & \hspace{0.3cm} 0 &
\hspace{0.3cm} -i\,3m_{f}\,{\rho_{0}}\,\omega\,({c_{s}}_{\parallel}) \\
-i\,\omega  & \hspace{0.3cm} i\,({c_{s}}_{\perp})\,k_{x} & \hspace{0.3cm} i\,({c_{s}}_{\perp})\,k_{y}
& \hspace{0.3cm} i\,({c_{s}}_{\parallel})\,k_{z}
\end{array} \right)
\label{mqhdeosAmatrix}
\end{equation}
which yields the following dispersion relation:
\begin{equation}
\omega^{4}-\Bigg[({{\tilde{c}_{s}}})^{2}(\,{k_{x}}^{2}+\,
{k_{y}}^{2}+ \,{k_{z}}^{2})+\Bigg({\frac{B^{2}\,
{Q_{f}}^{2}}{m_{f}^{2}}}\Bigg) \,\Bigg]\omega^{2}
+\Bigg({\frac{B^{2}\,{Q_{f}}^{2}}{m_{f}^{2}}}\Bigg)
({{\tilde{c}_{s}}})^{2} \,{k_{z}}^{2}=0
\label{mqcddisprelA}
\end{equation}
where we identify the
``effective sound speed'' ${{\tilde{c}_{s}}}$,
\begin{equation}
({{\tilde{c}_{s}}})^{2} \equiv   {\frac{9\,{g_{h}}^{2}\,{\rho_{0}}}{8\,m_{f}\,{m_{G}}^{2}}}
\label{newsound}
\end{equation}
which depends on the features of the EOS.  We can solve eq. (\ref{mqcddisprelA}) obtaining $\omega$ and the phase and group
velocities, which become identical with
equations (\ref{nrroots}) to (\ref{nrzgroupv})  when we
set ${c_{s}}_{\perp}={c_{s}}_{\parallel}={{\tilde{c}_{s}}}$ in the latter.
We can then conclude that stability and causality are  satisfied in the
present case.

Let us look at the following particular cases:
\vskip0.5cm
\noindent

$(i)$  {\it No B field} ($B =0$): \,\,\, $\omega(\vec{k})=
\pm ({\tilde{c}_{s}}) |\vec{k}|$ \,\,\,\, and
\,\,\,\, $|\vec{{{v}_{p}}}|= {\tilde{c}_{s}}$

In this case we recover the  results found in \cite{we13}.
\vskip0.7cm
\noindent

$(ii)$ {\it Very strong  B field} ($|Q_{f}|B \, > \, \mu^{2} \, > \, m_{f}^{2}$): The dispersion relation is:
\begin{equation}
\omega^{4}-\Bigg[({{\tilde{c}_{s}}})^{2}|\vec{k}|^{2}+({\mathcal{V}}_{s})^{2} \,{k_{z}}^{2}+
\Bigg({\frac{B^{2}\,{Q_{f}}^{2}}{m_{f}^{2}}}\Bigg) \,\Bigg]\omega^{2}
+\Bigg({\frac{B^{2}\,{Q_{f}}^{2}}{m_{f}^{2}}}\Bigg)
({\mathcal{V}}_{s})^{2} \,{k_{z}}^{2}
+\Bigg({\frac{B^{2}\,{Q_{f}}^{2}}{m_{f}^{2}}}\Bigg)({{\tilde{c}_{s}}})^{2}\,{k_{z}}^{2}=0
\label{smqcddisprelA}
\end{equation}
where  $({\mathcal{V}}_s)^2$ is given by (\ref{newsoundstrongBdetails}).  The condition  (\ref{nrjapascondsa}) is written as:
\begin{equation}
\lim_{|\vec{k}| \to \infty} \Bigg| {\frac{\omega}{|\vec{k}|}}
\Bigg| = \lim_{|\vec{k}| \to \infty} |\vec{{{v}_{p}}}|
\cong \lim_{|\vec{k}| \to \infty} \sqrt{({{\tilde{c}_{s}}})^{2} +
{\frac{({\mathcal{V}}_{s})^{2}{k_{z}}^{2}}{|\vec{k}|^{2}}}}
\cong \sqrt{ ({\tilde{c}_{s}})^{2} +({\mathcal{V}}_{s})^{2}
\, \cos^{2}(\theta)}
\label{strongBmqcdjapascondsa}
\end{equation}
The same discussion made below Eq. (\ref{sBmitjapascondsaA})
applies here. Causality may be  satisfied for
appropriate choices of $g_{h}/m_{G}$ and $\theta$.

\section{Conclusions}

We have studied the effects of a constant magnetic field on the
propagation of waves in  non-relativistic cold and ideal quark matter.
Using the equations of non-relativistic ideal hydrodynamics in an
external magnetic field, we have derived the dispersion relation for
density and velocity perturbations. The magnetic field was included both
in the equation of state and in the equations of motion, where the term
of the Lorentz force was considered.  We have used three equations of
state: a generic non-relativistic one, the MIT bag model EOS and the
mQCD EOS. The anisotropy effects caused by the $B$ field were also
manifest in the parallel and perpendicular sound speeds.
We proved that the introduction of the magnetic field does not lead to
instabilities in the velocity and density waves. In the case of the
non-relativistic EOS  the propagation of these waves was found to respect
causality. As for the MIT and mQCD equations of state, we found situations
where the phase velocity might be larger than one. In particular, this
might happen for waves moving along the direction of the (very strong)
magnetic field.  In spite of its limitations,
our study could determine the situations in where we are ``safe'' and
where we might expect problems with instabilities and causality.

\begin{acknowledgments}

This work was partially supported by the Brazilian funding agencies CAPES,
CNPq and FAPESP (contract 2012/98445-4). We are  grateful to M. Oertel,
M. Strickland and F. E. Mendon\c{c}a da Silveira for enlightening
discussions.
\end{acknowledgments}

\end{document}